# Temperature effect on lattice and electronic structures of $WTe_2$ from first-principles study

Gang Liu[1,2], Huimei Liu[1], Jian Zhou[3,a)], and Xiangang Wan[1,4]

[1]*National Laboratory of Solid State Microstructures, College of Physics, Nanjing University, Nanjing 210093, China*

[2]*School of Physics and Engineering, Henan University of Science and Technology, Luoyang 471023, China*

[3]*National Laboratory of Solid State Microstructures and Department of Materials Science and Engineering, Nanjing University, Nanjing 210093, China*

[4]*Collaborative Innovation Center of Advanced Microstructures, Nanjing University, Nanjing210093, China.*

## ABSTRACT

Tungsten ditelluride ($WTe_2$) exhibits extremely large and unsaturated magnetoresistance (MR). Due to the large spatially extensions of Te-5p and W-5d orbitals, the electronic properties of $WTe_2$ are sensitive to the lattice structures, which can probably affect the strongly temperature dependent MR found in experiment. Based on first-principle calculations, we investigate the temperature effect on the lattice and electronic structures of $WTe_2$. Our numerical results show that the thermal expansion coefficients of $WTe_2$ are highly anisotropic and considerably large. However, the temperature (less than 300 K) has ignorable effect on the Fermi surface of $WTe_2$. Our theoretical results clarify that the thermal expansion is not the main reason of the temperature-induced rapid decrease of magnetoresistance.

---

a) Electronic mail: zhoujian@nju.edu.cn

## I. INTRODUCTION

In the recent years, as a new class of two-dimensional (2D) materials, transition metal dichalcogenides (TMDs) with the formula $MX_2$, where M is a transition metal element (Mo, W, Re and so on) and X is a chalcogen element (S, Se and Te), have received much attention due to their diverse properties and promising potential applications.[1-12] Recently, a non-magnetic tungsten ditelluride ($WTe_2$) has attracted great attention owing to the interesting physical properties,[13-19] especially the extremely large positive magnetoresistance (MR) without saturation even at extremely high applied magnetic fields of 60T[14] and predicted type-II Weyl semi-metallicity.[20] A perfect balance between the electron and hole populations had been proposed as the primary source of these novel and unwavering MR effects[14], which however is questioned by very recent experiment.[21] The effect of spin-texture on the MR had also been suggested[13]. Basically the electron-hole compensation is a semi-classical mechanism, thus it is very interesting to notice that the MR is highly sensitive to the variation of temperature. The rapid decrease of MR at high temperature is explained by the decrease of the carrier mobilities.[22] However, as we know, the temperature can also modify the lattice structure, meanwhile, Te-5p and W-5d orbitals are spatially extended, thus making it very sensitive to structural variations.[9,23] Therefore, to check whether the thermal expansion and the related band-structure change response for the highly sensitive temperature-dependence of MR is an interesting problem, which we address in the present work.

Due to lack of the experimental thermal expansion data of $WTe_2$, we firstly calculate the thermal expansion coefficients (TECs) of $WTe_2$ by first-principles calculations. Because of the low symmetry in the orthorhombic $WTe_2$, the direct search for the free energy minima within the quasi-harmonic approximation at different temperature[24] is very time-consuming. Therefore in this work, we have adopted the Grüneisen formalism[25-27] to calculate the thermal expansion of orthorhombic $WTe_2$ by first-principles, which not only can save a lot of computational time, but also can obtain many other fundamental quantities simultaneously, such as the Grüneisen parameters, elastic constants and so on.

## II. COMPUTATIONAL DETAILS

All the first-principles calculations are performed by using the Vienna ab-initio simulation package (VASP)[28-31] based on density functional theory. The exchange-correlation functional is Perdew-Burke-Ernzerhof[32,33] of generalized gradient approximation, and the interaction between the core electrons and the valence electrons is schemed through the projector augmented wave[34,35] method. The total energy convergence criterion is $10^{-6}$ eV and the plane-wave cutoff energy is 300 eV. A Monkhorst-Pack[36] k-mesh of 14×7×3 is used to sample the Brillouin zone in the structure optimization. During the structural relaxations, the force convergence for ions is $10^{-3}$ eV/Å. Taking into account the importance of van der Waals (vdW) forces, the vdW-DF of correlationfunctional[37,38] is used in our calculation. The phonon dispersions are obtained by using the Phonopy package[39] based on the supercell approach with finite displacement method. To obtain the convergent phonon property in the calculations of harmonic interatomic force constants, a 3×2×1 supercell containing 72 atoms is used. An 87×48×21 q-mesh is used in the Grüneisen parameters calculations. The Fermi surfaces of WTe$_2$ in the $k_z$=0 plane are also calculated by VASP code with spin-orbit coupling included and atomic positions optimized.

The thermal expansion of WTe$_2$ is calculated based on the Grüneisen theory.[25-27] As an anisotropic system, the component of thermal expansion tensor **α** can be described as:

$$\alpha_{ab} = \frac{1}{VN} \sum_{\mathbf{q},\lambda} C_{\mathbf{q},\lambda} \sum_{d,e} S_{abde} \gamma_{\mathbf{q},\lambda}^{de}, \qquad (1)$$

$$\gamma_{\mathbf{q},\lambda}^{de} = -\partial \ln \omega_{\mathbf{q},\lambda} / \partial \epsilon_{de}, \qquad (2)$$

$$C_{\mathbf{q},\lambda} = \frac{(\hbar \omega_{\mathbf{q},\lambda})^2}{k_B T^2} \cdot \frac{\exp(\hbar \omega_{\mathbf{q},\lambda} / k_B T)}{\left[\exp(\hbar \omega_{\mathbf{q},\lambda} / k_B T) - 1\right]^2}, \qquad (3)$$

where $V$ is the volume of the unit cell, $N$ is the total number of **q** points in the first Brillouin zone, $\lambda$ is the index of the phonon modes and runs over all the phonon branches, $\epsilon_{de}$ and $S_{abde}$ are strain tensor and elastic compliance tensor, respectively. The elastic compliance tensor $S$ can be obtained by inverting the elastic matrix. $\gamma_{\mathbf{q},\lambda}^{de}$ is the generalized mode Grüneisen parameter,[40] and $C_{\mathbf{q},\lambda}$ is the contribution of phonon mode to the specific heat.

Due to the symmetry of $\mathbf{S}$, and the fact that $\gamma_{\mathbf{q},\lambda}^{de}=0$ when $d \neq e$,[40] we can express the linear thermal expansion coefficients of an orthorhombic system as follows with the Voigt's notation:[40,41]

$$\alpha_1 = \frac{1}{VN}\sum_{\mathbf{q},\lambda}C_{\mathbf{q},\lambda}(S_{11}\gamma_{\mathbf{q},\lambda}^1 + S_{12}\gamma_{\mathbf{q},\lambda}^2 + S_{13}\gamma_{\mathbf{q},\lambda}^3), \qquad (4)$$

$$\alpha_2 = \frac{1}{VN}\sum_{\mathbf{q},\lambda}C_{\mathbf{q},\lambda}(S_{12}\gamma_{\mathbf{q},\lambda}^1 + S_{22}\gamma_{\mathbf{q},\lambda}^2 + S_{23}\gamma_{\mathbf{q},\lambda}^3), \qquad (5)$$

$$\alpha_3 = \frac{1}{VN}\sum_{\mathbf{q},\lambda}C_{\mathbf{q},\lambda}(S_{13}\gamma_{\mathbf{q},\lambda}^1 + S_{23}\gamma_{\mathbf{q},\lambda}^2 + S_{33}\gamma_{\mathbf{q},\lambda}^3), \qquad (6)$$

where the LTECs along $a$, $b$, and $c$ directions are denoted by $\alpha_1$, $\alpha_2$, and $\alpha_3$, while $\gamma_{\mathbf{q},\lambda}^1$, $\gamma_{\mathbf{q},\lambda}^2$, and $\gamma_{\mathbf{q},\lambda}^3$ mean the generalized mode Grüneisen parameters along $a$, $b$, and $c$ directions, respectively.

## III. RESULTS AND DISCUSSIONS

WTe$_2$ is a layered TMD material with space group of *Pnm*2$_1$.[42] The structure is shown in Fig 1. The W atoms form zigzag metal-metal chains along the *a*-axis. The weak vdW forces dominate interactions between the Te-W-Te sandwich layers. The calculated lattice constants are listed in Table I, while the experimental values[42] are listed either for comparison. It is obvious that our calculated lattice constants are well consistent with the experimental values.

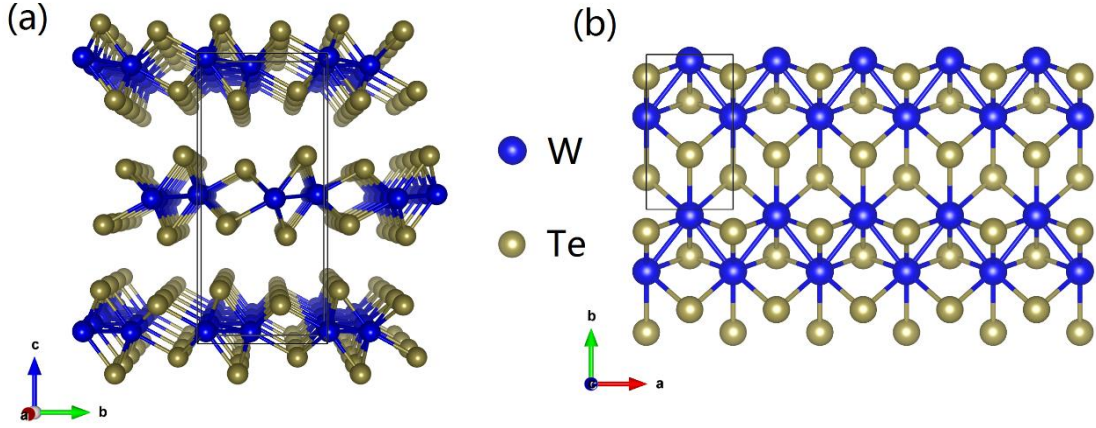

FIG. 1. Crystal structure of the bulk WTe$_2$ shown along the a-axis (a) and c-axis (b). Unit cell is shown in solid line.

We first calculated the generalized mode Grüneisen parameters by the central difference method along *a*, *b*, and *c* directions respectively. Then the so called density of generalized mode Grüneisen parameters defined by $g_i(\gamma) = \frac{1}{N}\sum_{\mathbf{q},\lambda}\delta(\gamma - \gamma^i_{\mathbf{q},\lambda})$[41] along three directions (*i*=1, 2, and 3 for *a*, *b*, and *c* respectively) is calculated and plotted in Fig. 2(a), while their detailed behaviors at low frequency is shown in Fig. 2(b). We can find that the density of Grüneisen parameters along *a* direction is always positive. But the negative value along *b* direction appears in very low frequency range below 15 cm$^{-1}$, and also the negative value along *c* direction appears in a higher frequency range of about 75~100 cm$^{-1}$. Since the positive mode Grüneisen parameters indicate a positive coefficient of thermal expansion,[43] we can speculate that the LTEC along *a* direction is always positive, while the LTEC along *b* direction may become negative at low temperature where only the low frequency acoustic modes are excited. However, there may not be any negative values of LTEC along *c* direction, since the negative density of Grüneisen parameters appears in the high frequency range. The phonon modes with negative Grüneisen parameters are only exited at high temperature, where more highly populated phonon modes at lower frequencies with positive Grüneisen parameters are also exited, canceling out the effect of these phonons.

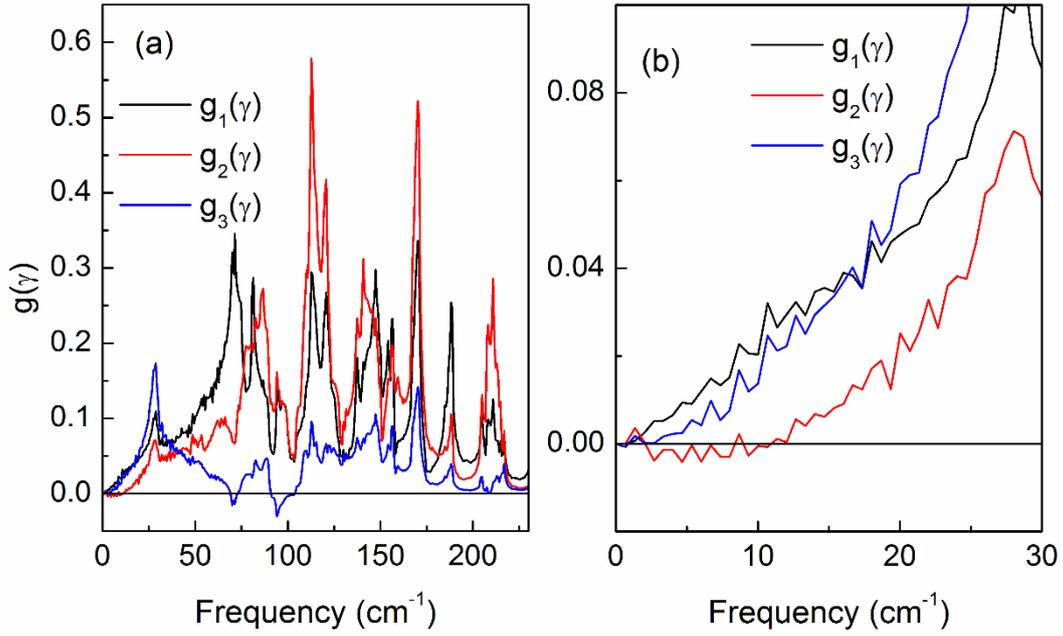

FIG. 2. (a) Frequency dependence of the density of generalized mode Grüneisen parameters. (b) The details of curves at low frequency.

Then, we calculate the temperature dependent LTECs and volumetric TEC of WTe$_2$ from 0 to 600 K, which is plotted in Fig. 3(a), while the detailed behavior at low temperature is shown in Fig. 3(b). We can find that LTECs of WTe$_2$ are significantly anisotropic in the whole range we investigated. The LTECs along *a* and *c* directions are always positive, and increase quickly as temperature increases. And the LTEC along *b* direction is slightly negative at low temperature. In Fig. 3(b), it can be found that the LTEC along *b*-axis is negative in the temperature range of 0~16 K, and it attains its lowest value of -7.65×10$^{-8}$ K$^{-1}$ at about 10 K. Then it rises quickly along with the increase of temperature. The LTECs along all the three directions reach plateaus when temperature is up to Debye temperature $\theta_D$= 137 K.[44]

At high temperature (much higher than $\theta_D$), the LTECs along *a*, *b*, and *c* directions attain their saturation values of 10.06, 7.54, and 4.45×10$^{-6}$ K$^{-1}$, respectively. We found that these values are at the same order compared with the recent experimental measurement of crystal Td-WTe$_2$.[45] However, it is interesting to find that our calculated LTEC along *c* direction (stacking direction) in WTe$_2$ is significantly smaller than those in *a* and *b* directions. But in the experiment, the LTEC along *c* direction is almost same as those in *a* and *b* directions.[45] Such a difference needs more further theoretical and experimental verification.

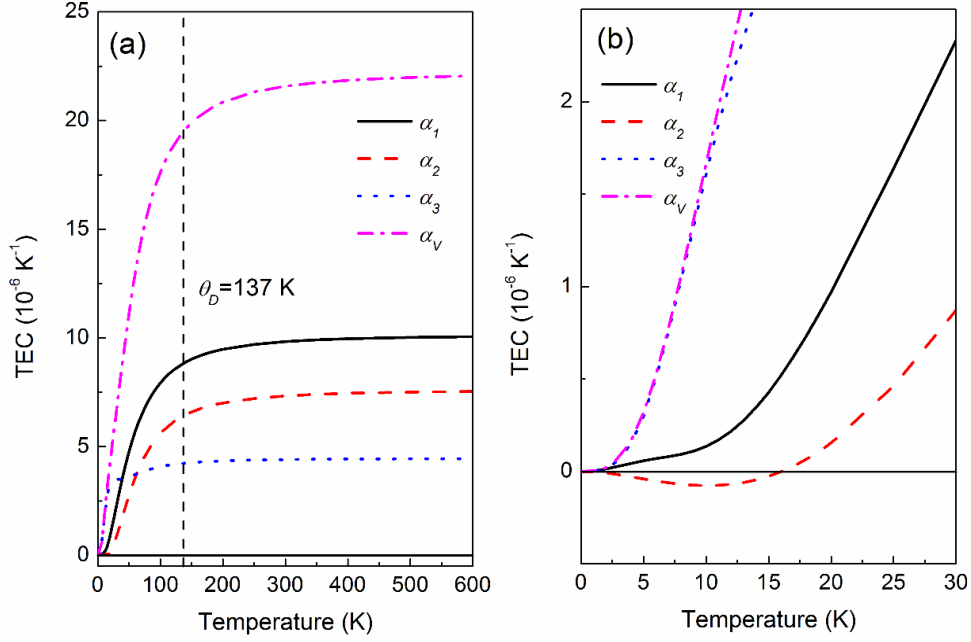

FIG. 3. (a) Temperature dependent linear and volumetric TECs of WTe$_2$. The details of curves at low temperature are displayed in (b).

We can also easily obtain the volumetric TEC $\alpha_V$ by using the formula $\alpha_V = \frac{1}{V}\left(\frac{\partial V}{\partial T}\right)_P = \alpha_1 + \alpha_2 + \alpha_3$, which is also plotted in Fig. 3. It can be found the volumetric TEC is always positive even at low temperature. This is because the absolute value of negative thermal expansion along *b*-axis is too small and balanced out by the positive thermal expansion along *a* and *c* directions. The behavior of temperature dependent volumetric TEC is similar to the LTECs. It increases quickly as temperature rises at low temperature, and then attains a plateau when temperature is up to $\theta_D$. At high temperature, the volumetric TEC reaches the saturation value of 22.05 ×10$^{-6}$ K$^{-1}$, which is also at the same order compared with the experimental results in Td-WTe$_2$.[45]

TABLE I. Calculated zero-temperature and other temperatures lattice constants of bulk WTe$_2$. Experimental ones (at 113 K)[42] are listed for comparison.

| Lattice constants (Å) | 0 K | 100 K | 200 K | 300 K | Exp. (113 K) |
|---|---|---|---|---|---|
| $a$ | 3.500 | 3.501 | 3.504 | 3.508 | 3.477 |
| $b$ | 6.289 | 6.291 | 6.295 | 6.300 | 6.249 |
| $c$ | 14.132 | 14.137 | 14.143 | 14.149 | 14.018 |

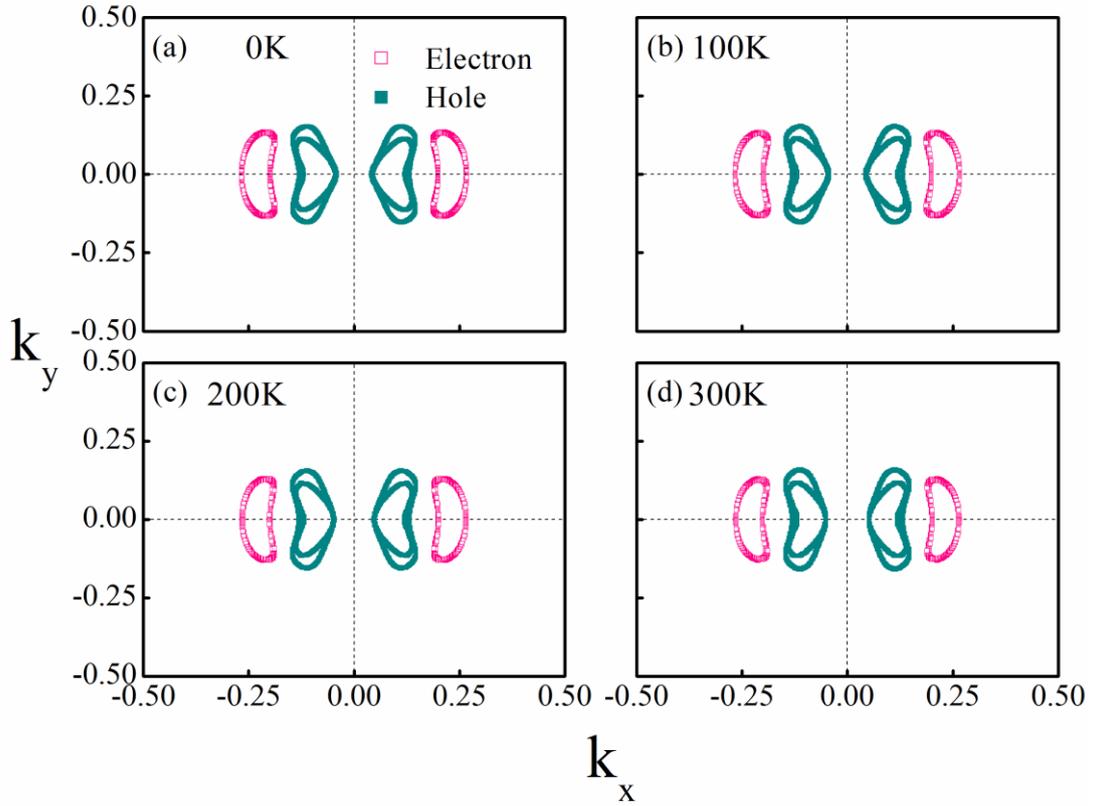

FIG. 4. Electron (red) and hole (green) Fermi surfaces from 0 to 300 K in $k_z=0$ plane.

Based on the above TEC, we then can easily obtain the lattice constants of WTe$_2$ at various temperatures, which is also given in Table I. Although the TEC of WTe$_2$ is considerably large, the lattice constants change little (less than 0.3% at most) in the range of 0-300 K. Finally, the fermi surfaces of WTe$_2$ at different temperatures can be calculated by taking different lattice constants from Table I. We first show electron and hole Fermi surfaces at $k_z=0$ plane in the Brillouin zone at 0 K in Fig. 4(a). We can find that there are two electron[46] and two hole pockets in WTe$_2$ and their size is quite similar. This indicates that the good balance of electron and hole population, resulting in the large MR found in experiments.

In Fig. 4(a-d), we present the evolution of Fermi surfaces of WTe$_2$ in an increasing temperature. It is obvious to find that the changes of Fermi surfaces are quite small from 0 to 300 K, which can be understood from the fact that the temperature has little effect on the lattice constant from 0 to 300 K. Therefore, the 'resonance' condition of electron and hole can always be satisfied and the decrease of large MR at room temperature in WTe$_2$ is maybe mostly due to the decrease of carrier mobility.

## IV. CONCLUSIONS

We have performed first-principles calculations to study the temperature dependent thermal expansion coefficients and their effect on the Fermi surfaces in Td-WTe$_2$. We find that the linear thermal expansion coefficients of WTe$_2$ are anisotropic, and they reach the saturated values of 10.06, 7.54, 4.45×10$^{-6}$ K$^{-1}$ along *a*, *b*, and *c* directions respectively at room temperature. However, since the temperature effect on the lattice constant is quite small, the electron and hole Fermi surfaces change little from 0 to 300 K.

## ACKNOWLEDGEMENTS

This work was supported by the National Key Technologies R&D Program of China (2016YFA0201104), National Basic Research Program of China (2015CB659400) and National Natural Science Foundation of China (No. 11474150, 11525417 and 11374137), the Open Project Program of Key Laboratory of Polar Materials and Devices, MOE (Grant No. KFKT20130001), East China Normal

University. The use of the computational resources in the High Performance Computing Center of Nanjing University for this work is also acknowledged.